# Photoionization and the formation of dwarf galaxies

Thomas Quinn[1], Neal Katz[1], and George Efstathiou [2]

[1]*Department of Astronomy, University of Washington, Seattle, WA 98195*
[2]*Department of Physics, University of Oxford, Oxford, OX1 3RH*

**SUMMARY**

It has been argued that a UV photoionizing background radiation field suppresses the formation of dwarf galaxies, and may even inhibit the formation of larger galaxies. In order to test this, we present gas-dynamical simulations of the formation of small objects in a CDM universe with and without a photoionizing background. The objects are selected from a collisionless simulation at a redshift of 2.4, and rerun at higher resolution including the effects of gas dynamics and using a hierarchical grid of particles. Five objects, each with a circular speed of $46\,\mathrm{km\,s^{-1}}$ are simulated. The presence of the photoionizing background has only a small effect on the amount of gas that collapses in these objects, reducing the amount of cold collapsed gas by at most 30%. Analysis of the smaller objects found in the higher resolution simulation indicates that the photoionizing background only significantly affects the formation of objects with a virialized halo mass less than $10^9\,\mathrm{M}_\odot$ and circular speeds less than $23\,\mathrm{km\,s^{-1}}$. However, the ionization balance is greatly changed by the presence of the background radiation field. Typical lines of sight through the objects have 4 orders of magnitude less neutral hydrogen column density when the photoionizing background is included.

## 1 INTRODUCTION

Hierarchical clustering models have been moderately successful in determining the upper end of the mass distribution of galaxies (White & Rees 1978). However, most models would predict a vast number of "minihalos" (*i.e.* galaxies with circular speeds $\lesssim 100\,\mathrm{km\,s^{-1}}$), and some mechanism is needed to prevent most of the baryonic material in the Universe from collapsing into such objects. Supernova driven winds (Dekel & Silk 1986) may provide a mechanism to suppress the formation of dwarf galaxies. Another suppression mechanism that may be effective for dwarf galaxies involves the presence of a photoionizing UV background radiation field that provides a source of heating and reduces the cooling of the baryonic material (Rees 1986, Efstathiou 1992). Evidence for such a field is given by the lack of a Gunn-Peterson effect even at a redshift of 4 (Webb *et al.* 1992), and the proximity effect in the Ly$\alpha$ forest (Lu *et al.* 1991). The estimated values of the background radiation field range from $10^{-22}$ to $10^{-21}\,\mathrm{erg\,cm^{-2}\,sr^{-1}\,Hz^{-1}\,s^{-1}}$ at the Lyman edge. The lower limit may be too low owing to obscuration by dust (Fall & Pei, 1993)

Efstathiou (1992) has calculated the cooling curve of a primordial gas with no metals in the presence of a background radiation field and has made simple models of the collapse of gas clouds. The photoionizing background has two effects. First, it raises the equilibrium temperature of the gas to about $10^4 - 10^5 K$. This raises the gas pressure and suppresses the collapse of gas into halos with circular speeds less than $20 - 50\,\mathrm{km\,s^{-1}}$. Secondly, the photoionizing background can suppress the cooling rate by reducing the neutral hydrogen and helium fraction, thus making the cooling times sufficiently long such that even larger objects might not collapse over the age of the Universe. Furthermore, the increased cooling time may make it easier for supernovæ to drive gas out of these objects. The simple models suggest that there can be significant increases in the cooling time since the neutral hydrogen and singly ionized he-lium that dominate cooling at low temperatures is eliminated. However, quantifying this effect on the formation of low mass galaxies requires numerical simulations.

In order for the gas to be unstable against collapse in a dark matter halo, the circular speed of the halo must satisfy (Rees 1986)

$$v_{equ} \gtrsim \sqrt{2} v_s = 12.8 \left(\frac{T}{10^4\,\mathrm{K}}\right)^{1/2} \left(\frac{\mu}{m_p}\right)^{-1} \mathrm{km\,s^{-1}}, \quad (1)$$

where $v_s$, $T$, and $\mu$ are respectively, the sound speed, the temperature, and the mean molecular weight of the gas, and $m_p$ is the proton mass. In a spherical collapse model (Gunn and Gott, 1972) using the mass within an enclosed density contour of 170 times the background density, this corresponds to a halo mass

$$\begin{aligned} M_{equ} \gtrsim & 1.09 \times 10^9 \left(\frac{T}{10^4\,\mathrm{K}}\right)^{3/2} h_{50}^{-1} \\ & \times (1+z)^{-3/2} \left(\frac{\mu}{m_p}\right)^{-3/2} \mathrm{M}_\odot. \end{aligned} \quad (2)$$

$h_{50}$ is the Hubble constant in units of 50 km/s/Mpc. For a neutral gas of primordial abundance at $T = 10^4$ K this implies a circular speed of 12.5 $\mathrm{km\,s^{-1}}$ and a mass of $1.6 \times 10^8\,\mathrm{M}_\odot$ at $z = 2.4$, while for a fully ionized gas at $T = 10^5$ K, the circular speed is 53 $\mathrm{km\,s^{-1}}$ and the mass is $1.2 \times 10^9\,\mathrm{M}_\odot$. If the increase in the cooling time caused by photoionization has a significant effect, then the formation of objects even larger than this will also be suppressed.

In the simulations that follow, we attempt to test the accuracy of these estimates regarding the sizes of objects whose formation are suppressed by the presence of a background radiation field. We first describe our series of simulations and the results, and then discuss the implications of these results.

## 2 SIMULATION

To simulate the collapse of individual clouds in their cosmological context, we use the multi-step approach described in Katz *et al.* (1994). First, we evolve a low resolution dissipationless simulation of a box 10 Mpc on a side ($H_0 = 50\,\mathrm{km\,s^{-1}}$) in a $b = 2$ CDM universe with $128^3$ particles ($m_p = 3.3 \times 10^7\,\mathrm{M}_\odot$) to a redshift of 2.4 using the PPPM N-body technique (Efstathiou *et al.* 1985). We use this simulation to select objects of the desired mass using a friends of friends algorithm with a linking length of 0.11 times the mean interparticle separation, corresponding to an overdensity of 178.

These objects are then simulated again with hydrodynamics using a hierarchical grid of particles that provides high resolution in the object while capturing the tides of the cosmological context. The effective resolution of these simulations is $256^3$ with the high resolution region starting as a sphere of radius 500 comoving kpc containing 8700 collisionless particles and 8700 gas particles. A further 4000 more massive particles are used to represent the evolution of the fluctuations in the rest of the 10 Mpc cube. On this hierarchical grid we impose the same fluctuations as the low resolution simulation, and we add fluctuations to continue the CDM spectrum to smaller scales. In accordance with big bang nucleosynthesis (Walker *et al.* 1991), we choose $\Omega_b = 0.05$, giving a dark matter particle mass of $3.93 \times 10^6\,\mathrm{M}_\odot$ and a gas particle mass of $2.07 \times 10^5\,\mathrm{M}_\odot$. We use a gravitational softening length of 0.5 comoving kpc for the high resolution particles, and the heavier particles have correspondingly larger softenings. We evolve the simulations using a comoving, periodic version of TREESPH (Hernquist & Katz 1989; Katz, Weinberg, & Hernquist 1995) with $\theta = 0.7$ and $2.26 \times 10^6$ years for the largest timestep. Owing to the multiple timestep nature of TREESPH some particles are integrated with timesteps 16 times smaller than the largest timestep. We start the simulations at a redshift of 20, and include both radiative and Compton cooling of the gas.

To study the effects of a photoionizing background, we simulate each object twice: once with an initial gas temperature of $10^4$ K and no photoionizing background, and once with an initial temperature of $10^5$ K and a uniform photoionizing flux of the form $J(\nu) = J_{-21}(z) \times 10^{-21} (\nu_L/\nu)^\alpha\,\mathrm{erg\,cm^{-2}\,sr^{-1}\,Hz^{-1}\,s^{-1}}$, where $\nu_L$ is the Lyman limit, $\alpha$ is 1, and $J_{-21}(z)$ evolves as

$$J_{-21}(z) = \frac{10}{1 + [5/(1+z)]^4} \quad (3)$$

as in Efstathiou (1992). This radiation field is much stronger than any estimates of the observed radiation field. We choose such a strong background field to quantify the maximum possible effect. The high initial temperature is also likely to be higher than expected if the intergalactic medium is photoionized, although temperatures of $\sim 50{,}000$K can be achieved if the intergalactic medium is photoionized impulsively (Miralda-Escudé and Rees, 1994). The treatment of the photoionization is fully described in Katz, Weinberg & Hernquist (1995), and is similar to the approach of Vedel *et al.* (1994).

We simulate five separate objects using this technique. We choose objects that are the smallest mass we could reasonably resolve in the low resolution simulation, or about 200 particles. This corresponds to a mass of $7.71 \times 10^9\,\mathrm{M}_\odot$ or a circular speed of 45.9 $\mathrm{km\,s^{-1}}$ at $z = 2.4$. The gas in such an object will be stable against collapse only at very high temperatures; ( $7.7 \times 10^4$ K ) however, at the level of background radiation chosen for these simulations the gas can have equilibrium temperatures of this order and cooling times longer than the

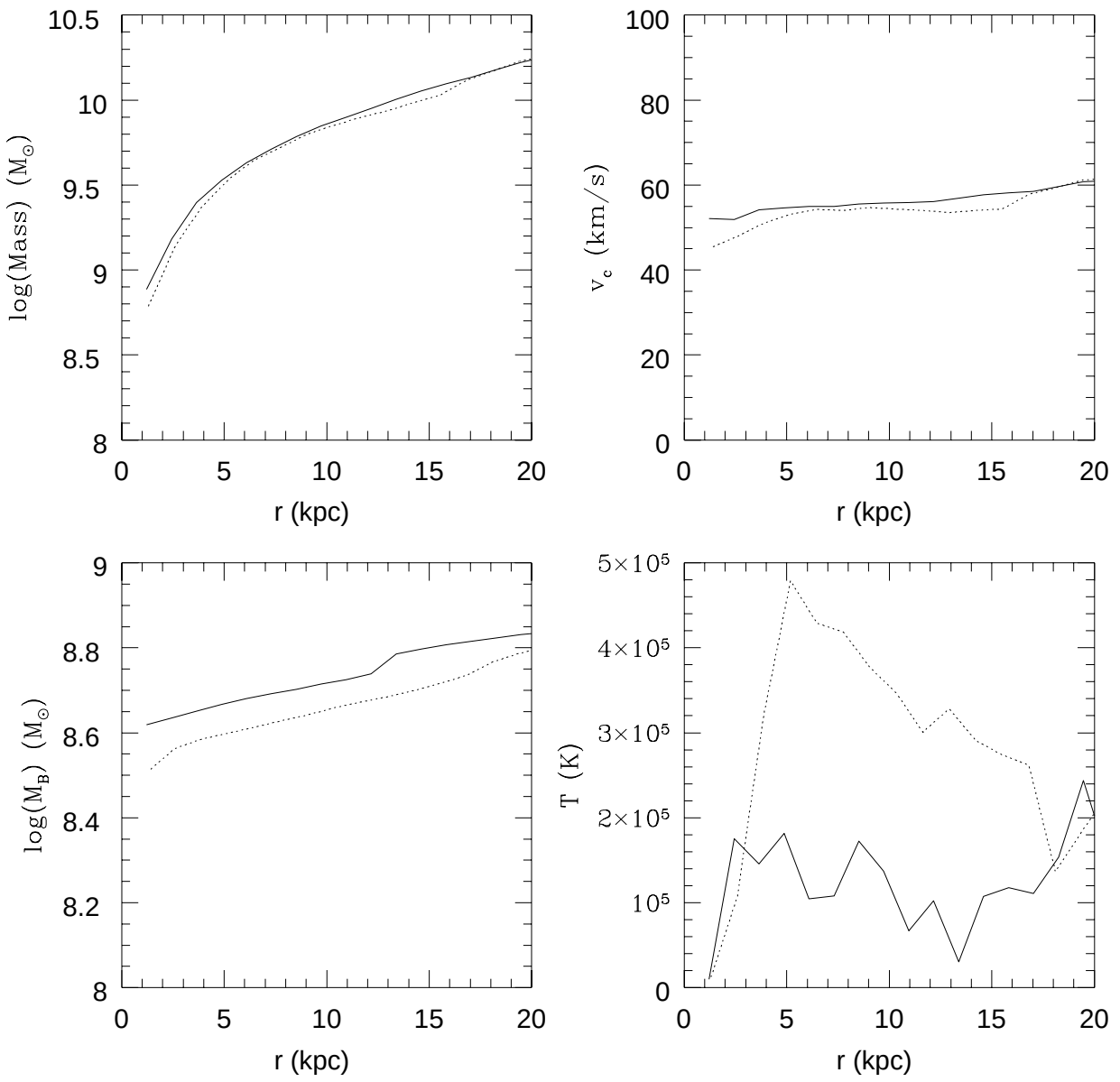

**Figure 1.** Comparison of the enclosed mass, circular speed, enclosed baryonic mass, and gas temperature as a function of radius for a halo at $z = 2.4$ with and without a photoionizing background. The solid lines are for the simulation without the photoionizing background, and the dotted lines are with the photoionizing background.

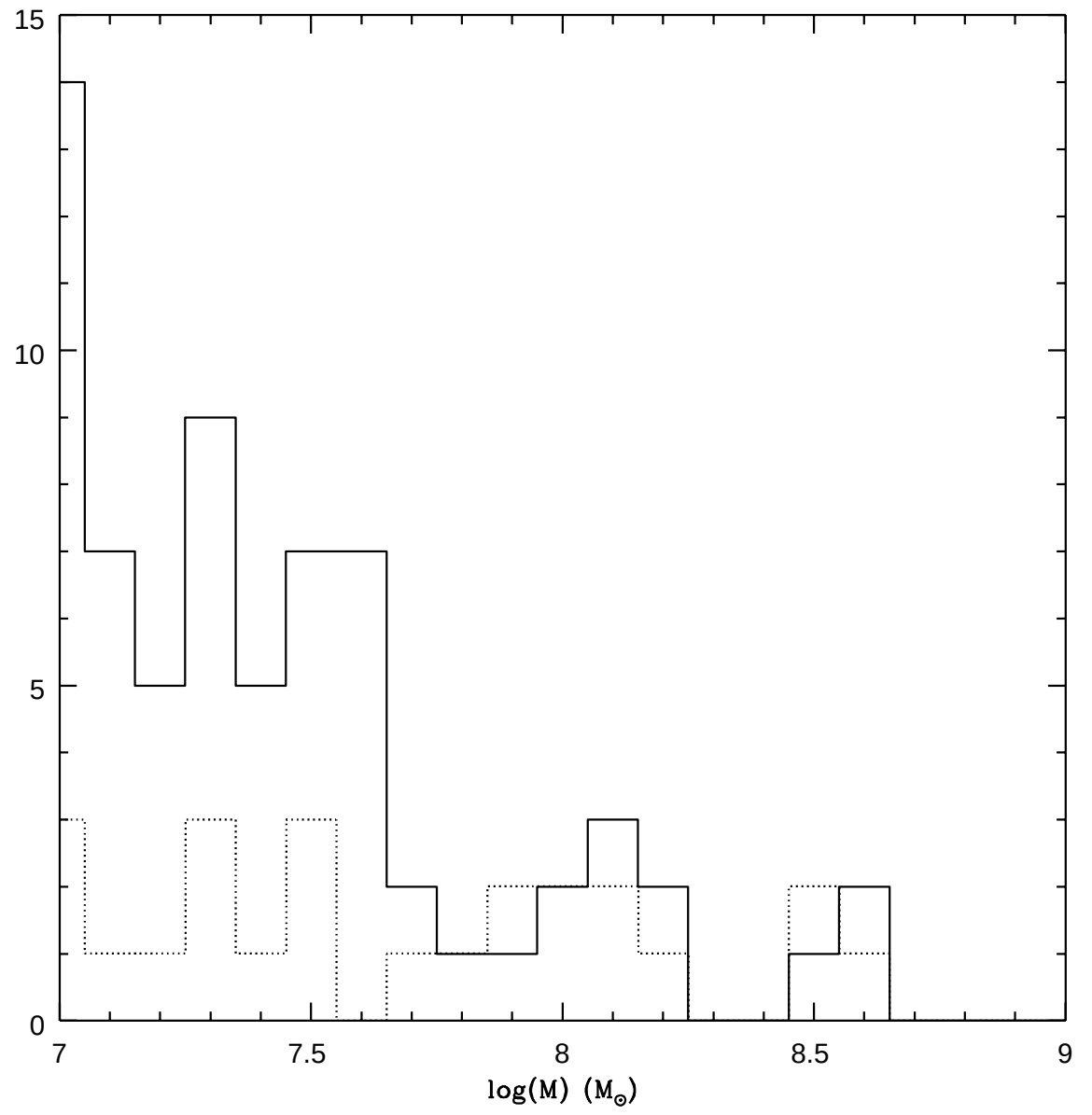

**Figure 2.** Distribution in the mass of cold collapsed gas for objects at $z = 2.4$. The solid lines are for the simulation without the photoionizing background, and the dotted lines are with the photoionizing background.

Hubble time at moderate ($\delta\rho/\rho \sim 10$–$100$) over densities. (See Efstathiou, 1992, Figure 2).

At $z = 2.4$ we compare the radial profiles of the chosen objects evolved with and without the photoionizing background. A typical comparison is shown in Figure 1, where we show the enclosed mass, the circular speed, the enclosed gas mass, and the gas temperature as a function of radius for one of our objects. The effect of the photoionizing background on the mass structure of this object is small. The total mass differs very little, and the gas mass is lowered by only 30%. However, the gas temperature outside the cold core is raised from $\approx 1.5 \times 10^5$ K in the no ionization case to nearly $5 \times 10^5$ K with the ionizing background. Of the five halos simulated, the one plotted has the greatest difference in the mass of gas with temperature less than $1.5 \times 10^4$ K between the runs with and without photoionization. Also note that for the run without photoionization 45% of the gas within the virial radius is in a diffuse hot component, while the run with photoionization has 54% of its gas in a hot component.

It appears that the UV background field has little effect on the collapse of gas into halos of the size that we chose to simulate in the low resolution simulation. However, our high resolution simulations include some halos that were too small to be well resolved in the low resolution simulation. Comparing these halos, we may be able to determine the mass at which the photoionizing background plays a significant role in the suppression of gas collapse. To find these halos, we use an algorithm similar to DENMAX (Bertschinger & Gelb, 1991; Stadel *et al.* 1995, in preparation) to identify bound groups of particles associated with density maxima. For each object we then determine how much of its mass is in cold ($T < 1.5 \times 10^4$ K) gas.

In Figure 2 we compare the distribution in the cold gas mass of the objects found in all the high resolution runs with and without the photoionizing background. The presence of the background field has only a small effect on the distribution of collapsed gas mass down to a total gas mass of $5 \times 10^7\,M_\odot$, but below this mass, the effect is quite marked. There are 54 objects with a cold gas mass between $1 \times 10^7\,M_\odot$ (50 particles) and $5 \times 10^7\,M_\odot$ (250 particles) in the simulations without the background field, while there are only 13 objects in this mass range in the simulations with the background field. Suppression of formation occurs below a total halo mass of $10^9\,M_\odot$ (a cold gas mass of $5 \times 10^7\,M_\odot$), or a circular speed of $23\,\mathrm{km\,s^{-1}}$. Another possibility that can't be completely ruled out is that we are reaching the resolution limit of our simulations (see Weinberg *et al.* 1995).

## 3 DISCUSSION

The simulations presented here suffer from many limitations, but these limitations arise from compromises made to attain as high a resolution as possible using a moderate number of particles. First, they were run to a fairly high redshift with a somewhat low amplitude of fluctuations. This gives the advantages of need-ing a smaller high resolution region, a smaller number of timesteps, and not going as far into the non-linear regime. Second, the dissipationless run from which we chose our objects is by no means a fair sample of the Universe, although this is ameliorated by the low amplitude fluctuations and high redshift. Thirdly, the high resolution region is just large enough to contain the material that fell into the central object. In fact, as many as 10 lower resolution particles with 8 times the mass of the high resolution particles fell within the virial radius of the central object at the end of the run. As well as their effect on the dynamics, the presence of these heavier particles indicates that more gas would have fallen into the object than we have accounted for, since only the high resolution region contains gas particles. Despite these drawbacks, these simulations are unique in their combination of a mass resolution of $2.05 \times 10^5\,\mathrm{M}_\odot$ with the representation of the cosmological context in a 10 Mpc cube.

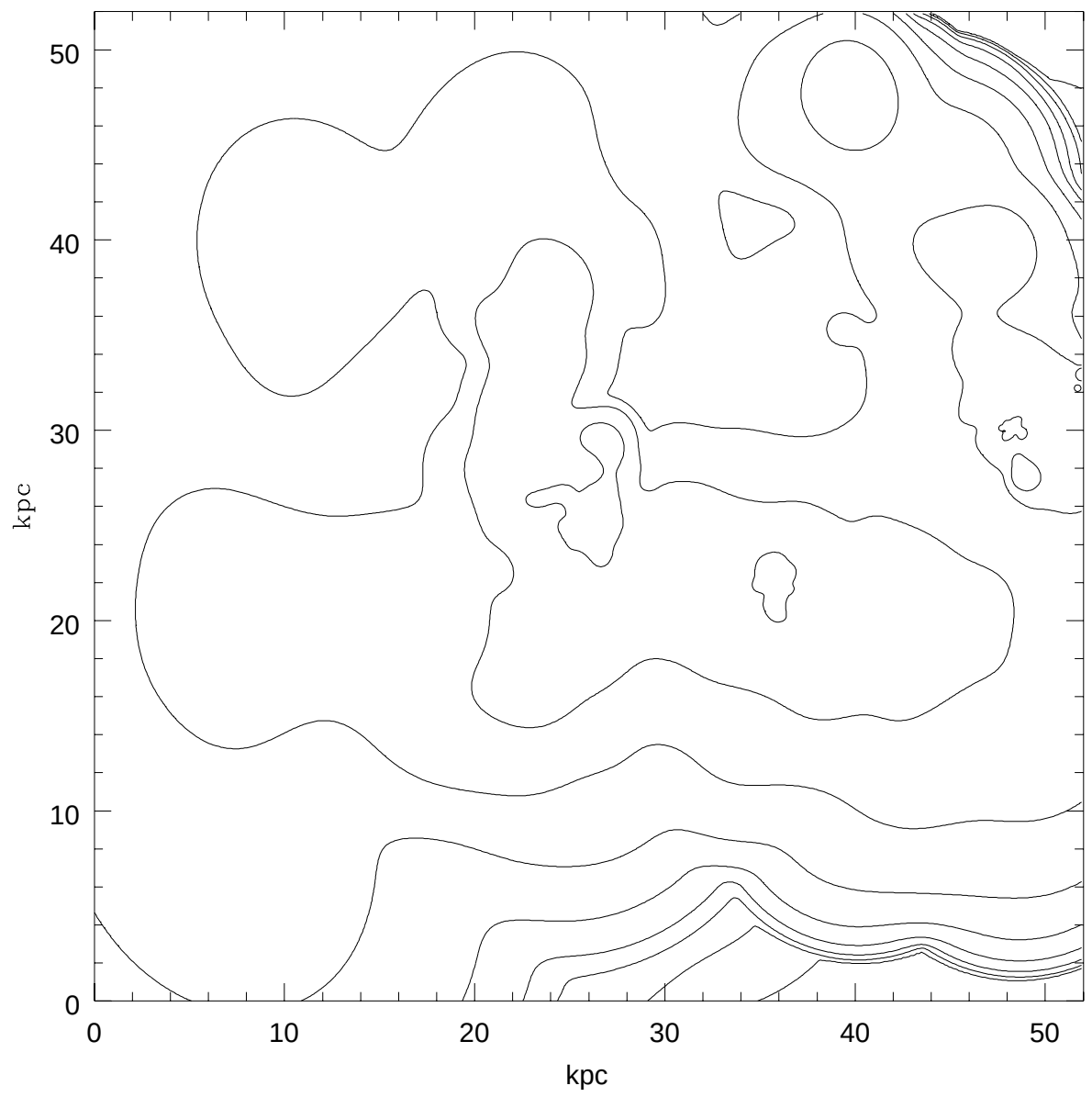

**Figure 3a.** Neutral Hydrogen map of a 52 kpc cube centered on an object simulated without a photoionizing background. The contours show the logarithm of the neutral hydrogen column density in atoms per square centimeter. They are spaced by factors of 10 from $10^{12}$ to $10^{20}$.

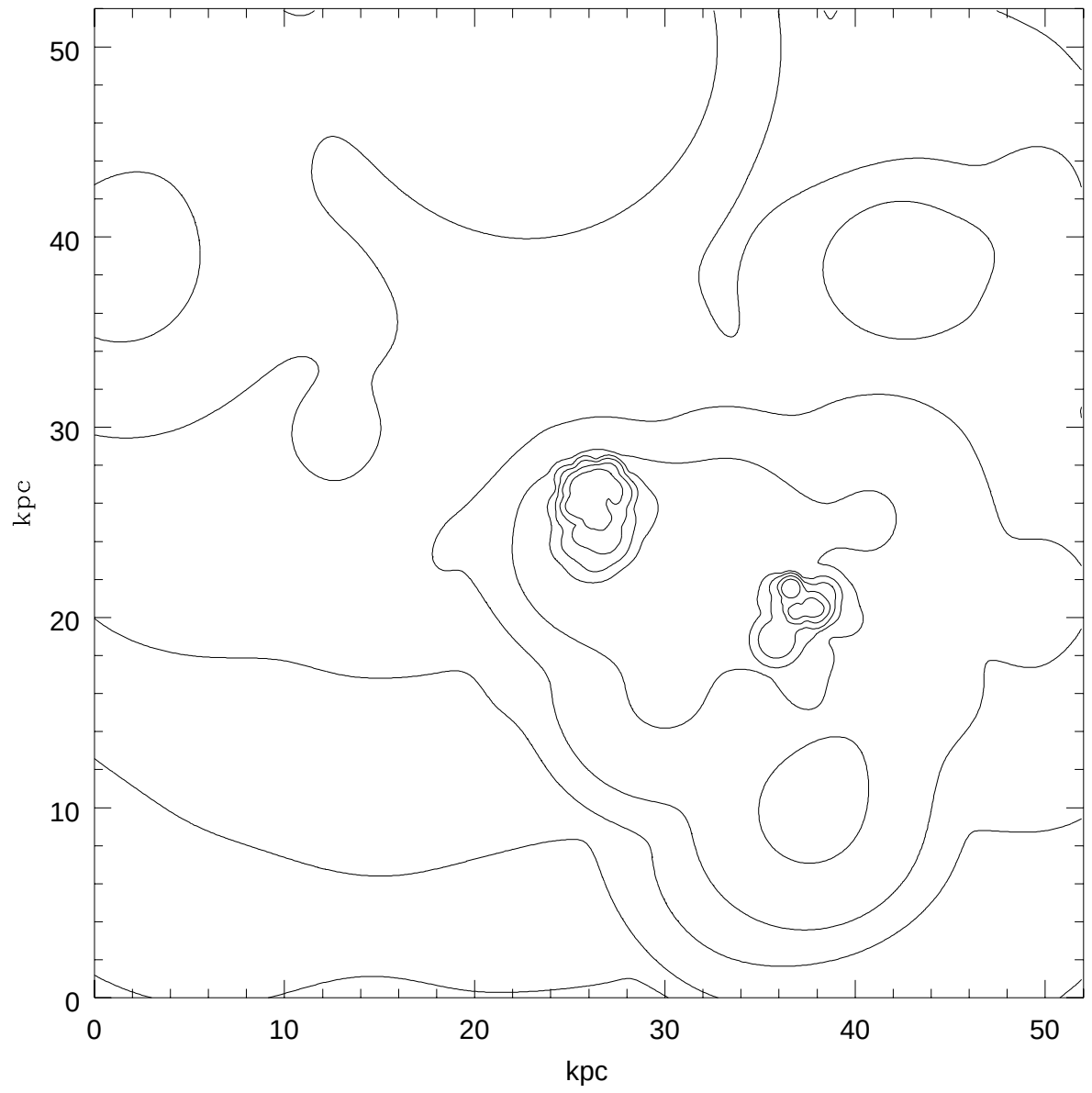

**Figure 3b.** Neutral Hydrogen map of the same object as figure 3a, but simulated with a photoionizing background. The contours have the same values as in figure 3a.

These simulations help to quantify some numbers in the analytic arguments of Efstathiou (1992). The equilibrium temperatures and particularly the cooling times are both density dependent. In the haloes of the objects the density is low enough that the photoionization has a significant effect. However, a significant amount of gas collapses to a high enough density that the suppression of the cooling rate by photoionization becomes ineffective. This conclusion is in rough agreement with the one dimensional hydrodynamic simulations of Thoul and Weinberg (in preparation). They find that a photoionizing background reduces the amount of cold collapsed gas for $50\,\mathrm{km\,s^{-1}}$ objects and completely suppresses the formation of objects with circular speeds less than $30\,\mathrm{km\,s^{-1}}$. The conclusions are also consistent with those of Weinberg, Hernquist, & Katz (1995) who conclude that a photoionizing background field does not affect the formation of objects with circular speeds $\sim 100\,\mathrm{km\,s^{-1}}$.

Our results show that photoionization alone is not sufficient to suppress the formation of dwarf galaxies in order to bring the predictions of hierarchical clustering models into agreement with observations. The hierarchical models tend to overproduce galaxies with luminosities less than $10^{10} L_\odot$ or circular speeds less than $100\,\mathrm{km\,s^{-1}}$. (White & Frenk, 1991) However, even with the generous amount of photoionizing flux in our simulations, we are not able to significantly suppress the formation of objects with circular speeds of $50\,\mathrm{km\,s^{-1}}$. Furthermore, most of the limitations of our simulations (lack of metal cooling, limited resolution, and missing infalling gas) make it harder to form objects. These results differ from those of Cen & Ostriker (1992) who find a good match to the observed galaxy luminosity function with a reasonable mass to light ratio. We believe that this discrepancy is due to different resolutions. Cen and Ostriker's highest resolution simulation has a resolution (two grid cells) of 30 kpc, a factor of 60 larger than our resolution. A lower resolution simulation is not able to resolve the small dense knots that cool quickly. The lower resolution simulations of Cen and Ostriker support this hypothesis, since their next lowest resolution simulation with 120 kpc resolution has 10 times fewer collapsed objects with $10^9\,M_\odot$ in baryons.

Although the UV background field only moderately affects the formation of $50\,\mathrm{km\,s^{-1}}$ objects, it completely changes their ionization structure, and therefore has significant implications for their absorption line characteristics. Figure 3a shows a map of the neutral hydrogen column density in atoms per square centimeter in a 52 kpc cube around an object simulated with no photoionizing background. Figure 3b shows the same object, but simulated with a photoionizing background. Note that over much of the area the neutral hydrogen column density is reduced from $10^{19}$ atoms/cm$^2$ to $10^{15}$ atoms/cm$^2$. The size of the region with greater than $10^{15}$ atoms/cm$^2$ is comparable to the size of collapsed objects in the simulation of Cen *et al.* (1994); however, this object has a high density knot not seen in the Cen *et al.* simulation, probably owing to their poorer spatial resolution. Since the simulations and the maps assume that the gas is optically thin to the UV radiation field, the column densities in Figure 3*b* are underestimated when the column densities are greater than about $10^{17}$ atoms/cm$^2$ (Katz, Weinberg, Hernquist & Miralda-Escudé, 1995). In a future paper, we will use the absorption cross-sections determined from simulations like these to calculate number densities of objects from the absorbers seen in quasar spectra.

## 4 ACKNOWLEDGMENTS

N. Katz and T. Quinn wish to acknowledge support from a NASA HPCC/ESS grant NAG 5-2213. N. Katz also received support from NASA Astrophysics Theory grant NAGW-2523.